\documentclass[aps,pra,reprint,showkeys]{revtex4-2}

\usepackage{amsmath}
\usepackage{graphicx}
\begin{document}

\title{A new approach to analyzing the spinor wave functions: \\
Effect of strain on the electronic structure and optical transitions in bulk CdSe \\}

\author{A. I. Lebedev}
\email[]{swan@scon155.phys.msu.ru}
\affiliation{Physics Department, Lomonosov Moscow State University}

\date{\today}

\begin{abstract}
An approach to analyzing the spinor wave functions that appear in the electronic structure
calculations when taking the spin-orbit interaction into account is developed. It is
based on the projection analysis of angular parts of wave functions onto irreducible
representations of the point group and analysis of the evolution of the energy levels
upon the adiabatic turning on the spin-orbit interaction. The technique is illustrated
by an example of the changes in the valence band structure in strained bulk CdSe with
zinc-blende structure. An analysis of the character of mixing of various branches of the
valence band supports the Luttinger--Kohn model of the valence band. It is shown that the
above calculations complemented by the Zeeman splitting in a magnetic field make it
possible to unambiguously determine the polarization of all optical transitions. Using
the spinor wave functions, matrix elements of optical transitions between the valence
subbands and the conduction band are calculated.
\end{abstract}

% insert suggested keywords - APS authors don't need to do this

\maketitle

\section{Introduction}

Being one of the perturbation theory techniques, the ${\bf k \cdot p}$
method~\cite{SemicondSemimetals.1.Ch3,VoonWillatzen} found wide application in solid
state physics. Strictly speaking, the use of this method requires to know the
energies of an infinite number of bands and an infinite number of matrix elements.
L{\"o}wdin~\cite{JChemPhys.19.1396} proposed a mathematical trick that allowed to project
this problem onto a system with a finite number of bands. It is in this truncated form that
the ${\bf k \cdot p}$ method is usually used as an ``interpolation scheme'' to describe
the band structure of semiconductors in the vicinity of a given point of the Brillouin zone.
In fact, this method is \emph{semi-empirical} since it borrows all necessary parameters
from the experiment. To calculate the dispersion curves, there is no need to know the wave
functions; their knowledge becomes necessary when one needs to calculate the matrix elements
of optical transitions and the character of their optical polarization.

A detailed comparison of predictions obtained using the first-principles calculations and
the ${\bf k \cdot p}$ method showed that in respect to low-dimensional structures (they
will be the subject of our further research), the predictions of the ${\bf k} \cdot {\bf p}$
method are characterized by a large number of shortcomings described and analyzed in detail
in Refs.~\cite{PhysRevB.53.7949,PhysRevB.57.9971,ElectrochemSocProc.98-19.259,
JPhysChemB.102.6449,PhysRevB.54.11417}. First of all, these are errors in the prediction
of energies of all subbands except for those that were used as empirical parameters;
overestimation of the size effect; impossibility to predict \emph{a full set} of subbands
in the conduction band~\cite{PhysRevB.53.7949,PhysRevB.57.9971} as well in the valence
band~\cite{JPhysChemB.102.6449}, and to describe optical transitions involving all subbands
of the conduction band~\cite{JPhysChemB.102.6449}. The main cause of these errors is the
insufficient number of bands taken into account in the calculations, which leads to
significant errors in description of dispersion curves. A basis of eight (including spin)
bands is the most commonly used in these calculations. At the same time, as was shown in
Ref.~\cite{PhysRevB.54.11417}, it is necessary to take at least 30 bands (including spin)
into account for a qualitative description of the band dispersion in GaAs, and its
quantitative description requires about 150 bands. When modeling the low-dimensional
structures within the ${\bf k \cdot p}$ method, one more problem arises: their description
in the envelope function approximation, which is based on the assumption of smoothness of
this function, is formally impossible due to the appearance the Fourier component of the
potential that goes beyond the first Brillouin zone. Additional difficulties arise
when modeling heterostructures with a strong lattice mismatch. These problems are absent
in the first-principles calculations.

The aim of this work was to develop an approach to analyzing the spinor wave functions
that would allow one to determine the genesis of various bands in low-dimensional
structures. The problem is that due to the size quantization effects and strong
mixing of different branches of the electronic spectrum, the identification of
electronic states in the valence band of these structures becomes extremely
complicated. Since optical transitions that take into account their polarization are
determined by the total angular momentum $j$ and magnetic quantum number $m_j$ in the
initial and final states, it becomes necessary to determine the contributions of states
with different~$(j,~m_j)$ to the corresponding wave functions. The available programs
for an analysis of spinor wave functions (see~\cite{ComputPhysCommun.272.108226} and the
references therein) are usually limited to determining the symmetry of the wave
functions; this is absolutely insufficient for determining the genesis and identifying
bands.

Cadmium selenide CdSe was chosen as an object on which the proposed approach will be
tested as it is a direct-gap semiconductor and has interesting optical properties.
A considerable interest in CdSe now is associated with the study of nanoparticles of
various shapes made of this
material~\cite{NanocrystalQuantumDots,Materials.3.2260,ChemRev.2c00436}.

In the bulk form, cadmium selenide usually crystallizes in the wurtzite hexagonal structure
(space group $P6_3mc = C_{6v}^4$) while its cubic phase with the zinc-blende structure
(space group $F{\bar 4}3m = T_d^2$) is metastable. However, under certain synthesis
conditions, samples can be obtained in the zinc-blende structure. In particular,
nanoplatelets and quantum dots prepared by chemical route at low temperatures often
have this structure.

Calculations of the electronic structure of cubic CdSe and its optical properties in a wide
energy range were carried out using OPW, LCAO, empirical pseudopotential, and FP-LAPW
methods~\cite{PhysRev.179.740,PhysRevB.47.9449,PhysRevB.49.7262,JChemPhys.124.104707,
JAlloysComp.509.6737}. The ${\bf k} \cdot {\bf p}$ method was used to calculate the
electronic structure and optical matrix elements in CdSe
nanoplatelets~\cite{NatureMater.10.936,PhysRevB.95.165414,JApplPhys.119.143107}.

\section{Calculation technique}

Calculations of the electronic structure of bulk CdSe with the zinc-blende structure
with taking the spin-orbit interaction into account were carried out from first principles
within the density-functional theory using the ABINIT program~\cite{abinit3}. The local
density approximation (LDA) and PAW pseudopotentials~\cite{ComputMaterSci.81.446} were
used in the calculations. The plane wave cutoff energy was 30~Ha (816~eV). The unit cell
parameters and atomic positions were fully relaxed until the forces acting on the atoms
became less than $5 \cdot 10^{-6}$~Ha/Bohr (0.25~meV/{\AA}) while the accuracy of the
total energy calculation was better than 10$^{-10}$~Ha ($3 \cdot 10^{-9}$~eV). A number
of calculations were also performed using the HGH pseudopotentials~\cite{PhysRevB.58.3641}
with a cutoff energy of 60~Ha (1632~eV) and the ONCVPSP ones~\cite{PhysRevB.88.085117}
with a cutoff energy of 50~Ha (1360~eV).

In real low-dimensional structures---superlattices, nanoplatelets, and
nanoheterostructures grown in the [001] direction---the layers are strained, and so
their symmetry is tetragonal or lower. This is why we consider CdSe as an active
region of such structures and study the properties of CdSe biaxially stretched and
biaxially compressed in the $xy$ plane.

Since the spin-orbit interaction practically does not affect the geometry of structures,
the calculations of the equilibrium geometry were carried out without taking spin into
account. The band energies and spinor wave functions in the obtained structures were
calculated with the spin-orbit interaction turned on. To output the values of the matrix
elements of the momentum operator, small changes were added into the code of the ABINIT
program. To unambiguously separate optical transitions from the light holes and of split-off
holes to the conduction band in $\sigma$ and $\pi$ polarizations, a magnetic field was
applied parallel to the [001]~axis of the sample. The possibility of such calculations was
recently added into the program~\cite{PhysRevB.99.184404}. Since these calculations in
the ABINIT program are implemented only for norm-conserving pseudopotentials, the HGH and
ONCVPSP pseudopotentials were used in these calculations.

Using two components of the new approach, namely, (1) tracing the changes in the
energies of bands upon the ``adiabatic'' turning on the spin-orbit interaction and
(2) determining the genesis of the spinor wave functions by expanding their angular
parts in terms of irreducible representations of the point group, it becomes possible
to identify the bands, trace their evolution, and determine the degree of mixing of
various electronic states in them and their symmetry. As far as the author knows,
this approach to an analysis of the spinor wave functions have not been used before.

Since CdSe is a direct-gap semiconductor with the band extrema at the $\Gamma$~point of
the Brillouin zone, we will be interested in optical transitions exactly at this point
when analyzing its optical properties. Upon a tetragonal distortion of the structure along
the $z$~axis, the FCC Bravais lattice transforms into a body-centered tetragonal Bravais
lattice whose translation vectors are rotated by 45$^\circ$ in the~$xy$ plane with respect
to the axes of the FCC lattice. Therefore, the directions of the $\Delta$ and $\Sigma$
axes in the Brillouin zone are also interchanged with respect to their directions in the
cubic structure.

\section{Group-theoretical background}

As will be shown in Sec.~\ref{sec41}, the valence band of cubic CdSe is mainly composed
of the $p$ orbitals of Se (orbital angular momentum $l = 1$). They will be denoted by
$|X \rangle$, $|Y \rangle$, and $|Z \rangle$. Without taking spin into account, the valence
band at the $\Gamma$~point is threefold degenerate and is described by a vector irreducible
representation $T_2$~\cite{AltmannHerzig} ($F_2$ according to Bir and Pikus~\cite{BirPikus1972-transl}).
When moving away from the $\Gamma$~point along the $\Delta$ or $\Lambda$ axes, these bands
split into a nondegenerate band of
light carriers (with magnetic quantum number $m = 0$) and a doubly degenerate band of
heavy carriers (with $m = \pm 1$). As will be shown below, the conduction band originates
from the $s$~states of Se and Cd atoms ($l = 0)$, is nondegenerate, and has a symmetry of
$A_1$ ($\Gamma_1$).

When the spin-orbit interaction is turned on, the symmetry of wave functions is described by the
irreducible representations of the double group in which the operations of the $T_d$ point group
are supplemented by the transformations of the spin variable. In contrast to five irreducible
representations of the $T_d$ point group at the $\Gamma$~point, there are three irreducible
representations in the double group~\cite{AltmannHerzig,Koster1963}. They are direct products
of representations of the point group and the ${\cal D}_{1/2}$ representation of the rotation
group for particles with spin~1/2. In particular, $\Gamma_6 = {\cal D}_{1/2} \otimes A_1$,
and the direct product ${\cal D}_{1/2} \otimes T_2$ is reducible and splits into the sum
$\Gamma_8 \oplus \Gamma_7$.

Spin-orbit interaction causes a complex mixing of $|X \rangle$, $|Y \rangle$, and $|Z \rangle$
orbitals and results in the lifting of the threefold degeneracy of the valence bands at the
$\Gamma$~point; the bands are split according to their total angular momentum~$j$. The heavy
holes HH ($j = 3/2$, $m_j = \pm 3/2$) and light holes LH ($j = 3/2$, $m_j = \pm 1/2$), which
originate genetically from the doubly degenerate band of heavy carriers with $m = \pm 1$, are
shifted upwards, and the split-off SO band ($j = 1/2$, $m_j = \pm 1/2$) arising from the
nondegenerate band of light carriers with $m = 0$ shifts downwards. The symmetry of the LH
and HH bands which are doubly degenerate at the $\Gamma$~point is described by the $\Gamma_8$
irreducible representation of the double group, the symmetry of the SO band is described
by the $\Gamma_7$ representation, and that of the conduction band is described by the
$\Gamma_6$ representation.

The wave functions of the conduction and three valence bands at the $\Gamma$~point in a cubic
crystal have the form~\cite{PhysRev.97.869,PhysRev.100.580,JPhysChemSolids.1.249}:
\begin{widetext}
    \begin{equation}
    \begin{gathered}
    \phi_{c1} = | iS \!\downarrow\rangle;~~\phi_{c2} = | iS \!\uparrow\rangle;~~\phi_{hh1} = \frac{1}{\sqrt{2}} | (X + iY) \!\uparrow\rangle;~~\phi_{hh2} = \frac{1}{\sqrt{2}} | (X - iY) \!\downarrow\rangle; \\
    \phi_{lh1} = \frac{1}{\sqrt{6}} | (X - iY) \!\uparrow\rangle + \sqrt{\frac{2}{3}}\, | Z \!\downarrow\rangle;~~\phi_{lh2} = \frac{1}{\sqrt{6}} | (-X + iY) \!\downarrow\rangle + \sqrt{\frac{2}{3}}\, | Z \!\uparrow\rangle; \\
    \phi_{so1} = \frac{1}{\sqrt{3}} | (X - iY) \!\uparrow\rangle - \frac{1}{\sqrt{3}} | Z \!\downarrow\rangle;~~\phi_{so2} = \frac{1}{\sqrt{3}} | (-X + iY) \!\downarrow\rangle - \frac{1}{\sqrt{3}} | Z \!\uparrow\rangle
    \end{gathered}
    \label{eq1}
    \end{equation}
\end{widetext}
(when choosing the coefficients, we follow the paper~\cite{JPhysChemSolids.1.249}). When
moving away from the $\Gamma$~point, the degeneracy of the LH and HH bands is lifted.

Biaxial compression or tension of the crystals in the (001) plane lowers their symmetry
to $I{\bar 4}m2$ ($D_{2d}^9$) and results in a significant change of their electronic
structure. From the group-theoretical point of view, when the symmetry of a crystal is
reduced to $D_{2d}$, the number of irreducible representations at the $\Gamma$~point of
the point group remain the same (five), but their number in the double group is reduced
to two ($\Gamma_6$ and $\Gamma_7$)~\cite{AltmannHerzig,Koster1963}. The vector representation
$T_2$ of the point group of the cubic structure becomes reducible: $T_2 \to B_2 \oplus E$.
Representations of the double group  in the tetragonal structure are obtained as follows:
$\Gamma_6 = {\cal D}_{1/2} \otimes A_1$ (conduction band),
$\Gamma_7 = {\cal D}_{1/2} \otimes B_2$ (SO band). Direct product ${\cal D}_{1/2} \otimes E$
(light and heavy holes) turns out to be reducible and decomposes into the sum
$\Gamma_6 \oplus \Gamma_7$. For odd spinor functions originating from $p$ atomic orbitals,
the $\Gamma_6$ symmetry indicates the heavy holes, and the $\Gamma_7$ one indicates the
light holes~\cite{AltmannHerzig}. The reducibility of the representation means that under
a tetragonal distortion, the degeneracy of the HH and LH bands is lifted at the
$\Gamma$~point ($\Gamma_8 \to \Gamma_6 \oplus \Gamma_7$). The fact that the states of
the LH and SO bands in a strained crystal are described by the same representation
($\Gamma_7$) indicates the possibility of their mixing.

Below, when analyzing the symmetry of spinor wave functions, we will use the projection
analysis to determine the expansion coefficients of the angular parts of wave functions
in an orthogonal basis of irreducible representations of the point group. In a cubic
crystal, all 4th order axes are equivalent, and the projection analysis does not allow
to resolve the contributions of bands with different~$m$ at the $\Gamma$~point. Under
a tetragonal strain ($F{\bar 4}3m \to I{\bar 4}m2$), the $z$~direction becomes
distinguished, the degeneracy of the $m = 0$ and $m = \pm 1$ states is lifted,
and unambiguous determination of the fraction of $|Z \rangle$ states in the LH and SO
bands of the~$\Gamma_7$ symmetry using the projection analysis becomes possible.
Taking the one-to-one correspondence between $JM_J$ and $LS$ bases into account, this
allows us to unambiguously determine the contributions of the $j = 3/2$ and $j = 1/2$
states to the wave functions of interest. This turns out to be very important for
determining the matrix elements of optical transitions.

In a cubic semiconductor, in accordance with Eq.~(\ref{eq1}), the contributions of the
$|X \rangle$, $|Y \rangle$, and $|Z \rangle$ orbitals to the SO band are the same (this
is a kind of a ``cubic reference point'' that arises from the equivalence of three
directions of a cube), and the projector analysis program usually gives misleading
results for the $|X \pm iY \rangle$ and $|Z \rangle$ projections. Under a tetragonal
distortion, these projections become well defined. As will be shown below, the projection
analysis finds a strong change in proportions of these states in a strained structure,
and since it significantly affects the matrix elements of optical transitions, this makes
it possible to change the absorption and luminescence spectra of a semiconductor using
the elastic strain.

We note that until now the discussion of the strain effects on the electronic structure
of the valence band within the ${\bf k} \cdot {\bf p}$ method was usually limited to
consideration of the band energies and their
dispersions~\cite{BirPikus1972-transl,VoonWillatzen,PhysicsOptoelectronicDevices}. One
of a few papers that has discussed the changes in the wave functions and matrix elements
of optical transitions in strained crystals with zinc-blende structure, was the
paper~\cite{IEEEJQuantElectron.29.1344}.

\section{Results and their discussion}

\subsection{Determination of symmetry and genesis of spinor wave functions}
\label{sec41}

As was already noted, the symmetry of the spinor wave functions at the $\Gamma$~point
in tetragonal structures can be either $\Gamma_6$ or $\Gamma_7$. Determination of the
symmetry of wave functions and their genesis proved to be not an easy task. The fact is
that in programs for electronic structure calculations, only the spatial symmetry of wave
functions (given by the point group) is taken into account. This is because in the double
group there is always the ${\bar E}$ operator, which corresponds to a rotation of a crystal
by 360$^\circ$ and changes the sign of the wave function to opposite. So, if the spin
variable were taken into account, then the resulting wave function would always be
identically equal to zero.

To find the double group representation that describes the symmetry of a given wave
function, we analyzed the symmetry of each of its spinor components for ${\bf k} = 0$
averaged in the~$xy$ plane with respect to reflections in the plane passing through the
cadmium atom. The components of the wave functions were extracted from the array of wave
functions using the \texttt{cut3d} program included into the ABINIT software package.
Although the
$I{\bar 4}m2$ space group does not contain the indicated symmetry operation, the improper
rotation~$S_4$ creates such an operation for the plane-averaged wave function. Even functions
correspond to the $\Gamma_6$~symmetry, odd functions correspond to the $\Gamma_7$ one. The
obtained symmetries of wave functions were confirmed later by calculations of the matrix
elements of optical transitions between the valence subbands and the conduction band; they
completely agree with the results obtained using the \texttt{IrRep} program~\cite{ComputPhysCommun.272.108226}.

\begin{figure*}
\centering
\includegraphics{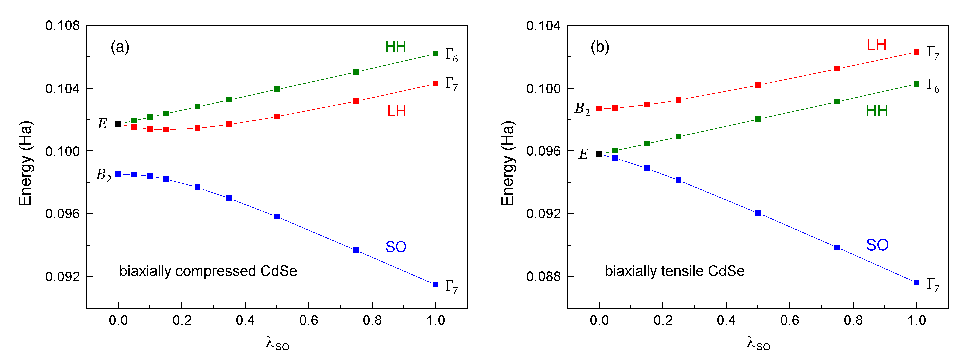}
\caption{\label{fig1}Changes in energies of the valence subbands at ${\bf k} = 0$ upon
the adiabatic turning on the spin-orbit interaction in (a) biaxially compressed and (b)
biaxially stretched CdSe crystals. The scaling parameter $\lambda_{\rm SO}$ denotes the
portion of the spin-orbit interaction energy used in the calculations. The amount of strain
is $\pm$1\%.}
\end{figure*}

As noted above, the existence of only two double group representations means that they
can arise from a superposition of four different states: light, heavy, split-off holes,
and $s$-like states of the conduction band in the initial cubic crystal. To determine
the genesis of spinor wave functions, the symmetry of both components of the spinor wave
function~$\psi$ for each band was analyzed in respect to all operations~$g$ of the point
group $G = D_{2d}$. The desired parameters were the values
    \begin{equation}
    F(g) = \frac{\| g \psi + \psi \|^2 - \| g \psi - \psi \|^2}{4 \| \psi \|^2},
    \label{eq2}
    \end{equation}
which were then decomposed using the character table in an orthogonal basis of irreducible
representations of the $D_{2d}$ point group. The decomposition coefficients
    \begin{equation}
    a_i = \frac{1}{| G |} \sum_{g \in G} \chi_i(g) \nu_i F(g),
    \end{equation}
where $| G |$ is the number of elements in the group, and $\chi_i$ is the character of
the $i$th representation, determine the contributions of different representations to the
analyzed wave function. The factor $\nu_i$ corrects the scalar product~(\ref{eq2}) for
the dimension of the $i$th representation. When performing the analysis, it turned out to
be important to choose the position of the reference point in the unit cell in such a way
that it coincided with one of the points of high symmetry; in our calculations, it was the
$2(a)$ Wyckoff position. To avoid the problem arising from the degeneracy of components of the
spinor wave function at ${\bf k} = 0$, the analysis was performed for the wave vector shifted
along the $\Sigma$~axis from the $\Gamma$~point to a distance of about 0.0005~{\AA}$^{-1}$.%
    \footnote{Due to the absence of an inversion center in crystals with the
    $D_{2d}^9$
    space group, the spinor wave functions are described by two independent components
    at all points of the Brillouin zone except for the $\Gamma$, $X$, $M$ points and the
    points on the~$\Lambda$ axis.}
As expected, the $A_1$, $B_2$, and $E$ representations appeared in the projections; the
projections onto $A_2$ and $B_1$ irreducible representations were absent.

An analysis of the projections of the obtained spinor wave functions onto atoms shows that
the conduction band of bulk CdSe is composed of 56\% of $s$~orbitals of Se atoms and 44\% of
$s$~orbitals of Cd atoms. The states in the valence band are formed mainly from $p$~orbitals
of Se with an admixture of about 7\% of $d$~states of Cd and 2\% $p$~states of Cd and
therefore, according to their genesis, can be classified as light, heavy, and split-off
holes. The main problem in the interpretation of these bands is how to distinguish them.
There is no problem for the heavy holes since their wave functions have the $| X \pm iY \rangle$
symmetry (\ref{eq1}), and so under the projection analysis they show zero $| Z \rangle$
component. For light and split-off holes the situation is more complicated, however
the study of the behavior of $| Z \rangle$ projections of wave functions on the scaling
parameter $\lambda_{\rm SO}$, which determines the degree in which the spin-orbit
interaction is taken into account in the calculations ($\lambda_{\rm SO} H_{\rm SO}$)
can help: with an increase in this parameter, the magnitude of projections should tend
to fractions of $| Z \rangle$ states predicted by formulas~(\ref{eq1}). Such calculations
can be performed within the ABINIT program, in which the scaling parameter $\lambda_{\rm SO}$
used in the calculations with PAW pseudopotentials is controlled by the \texttt{spnorbscl}
variable.

An analysis of the projections of spinor wave functions onto irreducible representations
of the point group shows that the symmetry of the conduction band (CB) in CdSe is closest
to the $s$ type: the angular part of its wave function $\Gamma_6$ is projected onto the
$A_1$ representation; the admixture of contributions of other symmetries does not exceed
0.1\%. The angular part of the wave functions of HH band with the $\Gamma_6$ symmetry is
projected onto the $E$~representation; an admixture of wave functions of other symmetries
does not exceed 0.1--0.2\%. Despite the same symmetry of the CB and HH bands, they have
different parity with respect to the inversion and, therefore, they do not mix at the
$\Gamma$~point (their mixing becomes possible at ${\bf k} \ne 0$). As expected, the
strongest mixing at the $\Gamma$~point is observed for the LH and SO bands, which have
the same $\Gamma_7$ symmetry. As the sum of fractions of the $| Z\rangle$ states in these
bands is close to unity, this indicates the weakness of interaction with remote bands and
the adequacy of the assumptions underlying the Luttinger--Kohn model of the valence
band~\cite{PhysRev.97.869}.

The evolution of the energies of various branches of the valence band upon the adiabatic
turning on the spin-orbit interaction in biaxially compressed and biaxially stretched
CdSe crystals is shown in Fig.~\ref{fig1}. When the spin-orbit interaction is turned off
($\lambda_{\rm SO} = 0$), there is already a splitting of bands caused by strain in crystals.
When the spin-orbit interaction is turned on, the sequence of bands in a biaxially compressed
sample corresponds to its usual order (the upper subband is the HH band, the following
subband is the LH one, and the lowest subband is the SO band). The surprising thing is
that at $\lambda_{\rm SO} = 0$ it is \emph{the SO band} that is split off: it is commonly
believed that the strain removes the degeneracy of HH--LH degenerate bands.

In biaxially stretched samples the situation is also surprising. The strangeness here
is in that for $\lambda_{\rm SO} \to 0$ the HH and SO branches, which have completely
different $j$ and $m_j$, turn out to be degenerate.

The ${\bf k \cdot p}$ method can help us understand the reason for what is happening.
In the $JM_J$ basis~(\ref{eq1}), by taking into account only the biaxial strain
$\epsilon_{xx} = \epsilon_{yy} = \epsilon$, the $\epsilon_{zz} = - 2\nu\epsilon/(1 - \nu)$
condition to get zero stress along the $z$ axis, the calculated Poisson's ratio for
bulk CdSe ($\nu = 0.422$), and neglecting the isotropic strain which simply shifts all
three bands by the same amount, the $6 \times 6$ block corresponding to the valence band
subbands in the Hamiltonian at ${\bf k} = 0$ is~\cite{VoonWillatzen}
    \begin{equation}
    H_\epsilon + H_{\rm SO} =
    \begin{vmatrix}
    H_1 & 0 \\
    0   & H_1 \\
    \end{vmatrix}
    ,~~H_1 =
    \begin{vmatrix}
    b_1\tilde{\epsilon} & 0                    & 0 \\
    0           & -b_1\tilde{\epsilon}         & \sqrt{2} b_1\tilde{\epsilon} \\
    0           & \sqrt{2} b_1\tilde{\epsilon} & -\lambda_{\rm SO} \Delta_{\rm SO} \\
    \end{vmatrix}
    ,
    \label{eq4}
    \end{equation}
where $\tilde{\epsilon} = \epsilon_{zz} - (\epsilon_{xx} + \epsilon_{yy})/2 = -2.460\epsilon$.
Here the rows and columns in the $3 \times 3$ matrix correspond to the HH, LH, and
SO bands,  respectively,  $b_1$ is the deformation potential of anisotropic strain,
$\epsilon$ is the magnitude of this strain, and $\Delta_{\rm SO}$ is the spin-orbit
splitting. At $\epsilon \ne 0$ and $\lambda_{\rm SO} = 0$, the bands form a singlet
with the energy $-2b_1\tilde{\epsilon}$ and the wave function $| Z \rangle$ (of
symmetry~$B_2$) and a doublet with the energy $+b_1\tilde{\epsilon}$ and wave function
$| X \pm iY \rangle / \sqrt{2}$ (of symmetry $E$). One of the wave functions of the
doublet and the wave function of a singlet turn out to be linear combinations of the
$j = 3/2$ and $j = 1/2$ states (the symmetry of the $D_{2d}$ group admits such a
mixing~\cite{AltmannHerzig}). Given that usually $b_1 < 0$~\cite{VoonWillatzen},
we get the energy spectrum shown in Fig.~\ref{fig1}.

\begin{figure}
\centering
\includegraphics{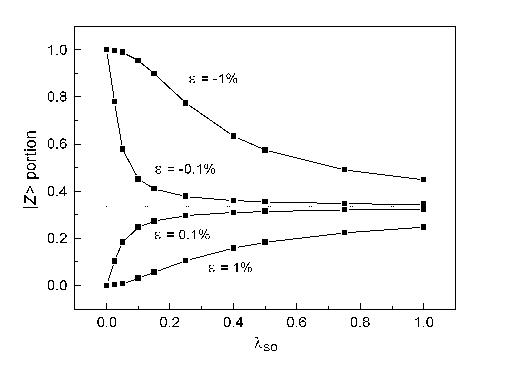}
\caption{\label{fig2}Fraction of $| Z \rangle$ states in the wave function of SO
subband as a function of the scaling parameter $\lambda_{\rm SO}$ for several values
of biaxial strain~$\epsilon$ of bulk CdSe. The dotted line shows the fraction of
$| Z \rangle$ states, at which the wave function becomes a true function corresponding
to $j = 1/2, m_j = \pm 1/2$.}
\end{figure}

When the spin-orbit interaction is turned on, the degeneracy of the doublet is removed,
and the proportion between the $j = 3/2$ and $j = 1/2$ components in wave functions
of the LH and SO subbands evolves to the values described by the equation~(\ref{eq1}):
as can be seen in Fig.~\ref{fig2}, when changing $\lambda_{\rm SO}$, the fraction of
$| Z \rangle$ projection in the SO wave function tends asymptotically to 1/3 for all values
of $\epsilon$. From Eq.~\ref{eq4} it follows that the proportion between the $j = 3/2$
and $j = 1/2$ components in the wave functions is controlled by the dimensionless parameter
$\lambda_{\rm SO} \Delta_{\rm SO}/b_1 \tilde{\epsilon}$.

\begin{figure}
\centering
\includegraphics{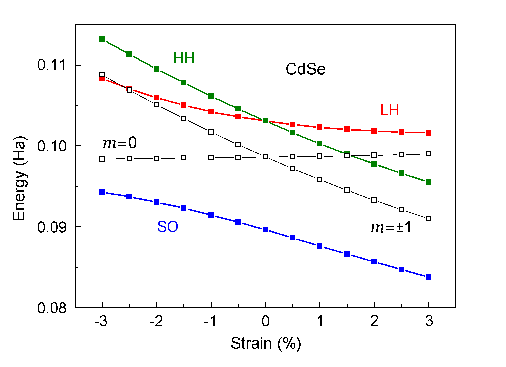}
\caption{\label{fig3}Effect of the biaxial strain on the position of valence subbands at
${\bf k} = 0$. Thin black lines are the band energies without the spin-orbit interaction,
color lines are the band energies when the spin-orbit interaction is taken into account.}
\end{figure}

The changes in positions of the valence subbands at ${\bf k} = 0$ as a function of strain
are shown in Fig.~\ref{fig3}. It is seen that without the spin-orbit interaction (thin
black lines) a tetragonal strain always results in splitting of states with $m = \pm 1$
and $m = 0$ (as noted above, the latter state is a predecessor of the split-off band).
When the spin-orbit interaction is turned on, the heavy holes band ($\Gamma_6$) exhibits
an upward shift parallel to the $m = \pm 1$ curve while the two remaining $\Gamma_7$ bands
undergo a complex interaction and mixing. The latter is evidenced by the observation of
an anticrossing pattern characteristic of interacting bands with a minimum gap between
the bands at a strain of $\approx -1.37$\%.

\begin{figure}
\centering
\includegraphics{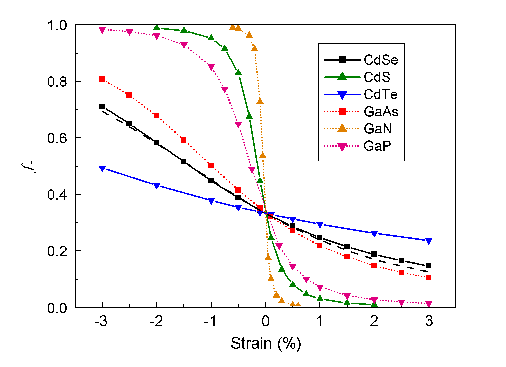}
\caption{\label{fig4}Fraction of $| Z \rangle$ states in the wave function of SO
subband as a function of biaxial strain for CdSe and several other bulk semiconductors
with zinc-blende structure. The dashed line shows the results of fitting the calculated
data for CdSe using Eq.~(\ref{eq5}).}
\end{figure}

The fraction of $|Z \rangle$ states in wave functions of the SO band as a function of
biaxial strain is shown in Fig.~\ref{fig4}. It is seen that the contribution of the
$|Z \rangle$ states into this band rapidly increases when the samples are compressed
and rapidly decreases when the samples are stretched. As follows from the figure, this
behavior is general for semiconductors with the zinc-blende structure. For the
Hamiltonian~(\ref{eq4}), the fraction of $|Z \rangle$ states in the wave function of
SO subband is described by the formula
    \begin{equation}
    f_z = \frac{1}{2} \Big(1 - \frac{1 - 3t}{\sqrt{9 - 6t + 9t^2}}\Big),
    \label{eq5}
    \end{equation}
where $t = 3 b_1 \tilde{\epsilon} /\lambda_{\rm SO} \Delta_{\rm SO}$ (this formula was
previously obtained in paper~\cite{IEEEJQuantElectron.29.1344} using different basis
functions). As follows from Fig.~\ref{fig4}, in which this dependence for CdSe is shown
by a dashed line, at the value of $b_1 = -1.38$~eV the results of first-principles
calculations agree well with predictions of the above formula. The obtained parameter
is in a reasonable agreement with the $-$1.1~eV value calculated for cubic CdSe within
the LCAO $sp^3s^*$ model in Ref.~\cite{PhysStatSolidiB.126.11}. We note that the minimum
gap between the LH and SO subbands appears at the strain value of $\approx -1.37$\%, at
which the mixing of these subbands is maximal (their proportion is 1:1).

\subsection{Matrix elements of optical transitions}

The group theory states that for optical transitions to be allowed in the dipole
approximation, the direct product of representations of groups describing the initial and
final states should contain at least one vector representation. For the $D_{2d}$ group
we have
$\Gamma_6 \otimes \Gamma_6 = A_1 \oplus A_2 \oplus E$, $\Gamma_6 \otimes \Gamma_7 = B_1 \oplus B_2 \oplus E$,
$\Gamma_7 \otimes \Gamma_7 = A_1 \oplus A_2 \oplus E$~\cite{AltmannHerzig,Koster1963}.
This means that in the crystals under discussion, the $\Gamma_6$--$\Gamma_7$ optical
transitions are allowed in both ($\sigma$ and $\pi$) polarizations, whereas the
$\Gamma_6$--$\Gamma_6$ and $\Gamma_7$--$\Gamma_7$ transitions are allowed only in the
$\sigma$~polarization, in which the electric field vector lies in the~$xy$ plane.

\begin{figure}
\centering
\includegraphics{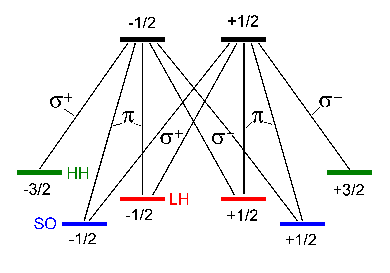}
\caption{\label{fig5}Polarization of light accompanying direct interband optical transitions
between states with different projections of the total angular momentum $m_j$~\cite{OpticalOrientation}.}
\end{figure}

By examining the scheme of possible optical transitions (Fig.~\ref{fig5}), one can see that
the only possible transition between the HH and CB bands is a transition in the $\sigma$
polarization with a change of projection of the total angular momentum of $\Delta m_j = \pm 1$.
The transition with $\Delta m_j = \pm 2$ is forbidden in the dipole approximation, and the
transition between the bands in the $\pi$ polarization ($\Delta m_j = 0$) are impossible
because of the lack of a suitable final state in the conduction band.

For the LH$\to$CB and SO$\to$CB transitions, the identification of transitions is complicated
by the equality $|m_j| = 1/2$ for all involved states and the spin degeneracy of them at
${\bf k} = 0$. Although the transitions in $\sigma$ and $\pi$ polarizations can be
distinguished by the values of matrix elements of the momentum operator $|P_x|^2$ and $|P_z|^2$,%
    \footnote{To do this, we need a tetragonal strain; without it, the matrix elements in
    all polarizations are the same (cubic symmetry).}
the spin degeneracy does not enable in our program to distinguish between the wave functions
with $m_j = 1/2$ and $m_j = -1/2$ and so to unambiguously identify all transitions.%
    \footnote{In particular, this peculiarity results in appearance of forbidden transitions
    with $\Delta m_j = 2$ when calculating the HH$\to$CB matrix elements with these wave
    functions.}
For their unambiguous
identification, we used a trick with the imposition of an external magnetic field, which
removes the degeneracy of these levels due to the Zeeman splitting. This allowed us to
solve our problem. The only unresolved question was that the sign of the $g$~factor in the
Zeeman splitting remains unknown. If we take its sign to be positive for electrons, then
the same sign applies for heavy holes, but the signs for other two bands are opposite to
each other and depend on the strain. At a strain close to zero, the sign of the $g$~factor
for the SO band is positive while for the LH band it is negative. However, under strong
biaxial compression, the signs of $g$~factors for these bands are reversed. Curiously, the
change in signs of $g$~factors for these two bands occurs at approximately the same strain
at which the contributions of the $| Z \rangle$ states into the bands become equal. The
results with imposition of a magnetic field obtained using the HGH pseudopotentials were
completely confirmed by calculations with norm-conserving ONCVPSP pseudopotentials~\cite{PhysRevB.88.085117}
(we note that in the used version of the ABINIT program (8.10.3) such calculations are
possible only for full-relativistic LDA pseudopotentials). The spin-resolved selection
rules obtained within the described approach are in full agreement with the group-theoretical
predictions.

\begin{figure}
\centering
\includegraphics{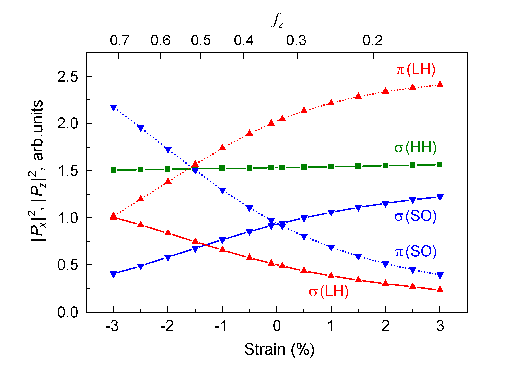}
\caption{\label{fig6}Squares of matrix elements of optical transitions $|P_x|^2$ and
$|P_z|^2$ as a function of biaxial strain of bulk CdSe. The top axis shows the
fraction of the $| Z \rangle$ states in the SO wave function.}
\end{figure}

The squares of the matrix elements of individual transitions in CdSe as a function of
strain are shown in Fig.~\ref{fig6}. At zero strain, the probabilities of individual
transitions completely correspond to the probabilities expected for cubic
crystals~\cite{OpticalOrientation}. At non-zero strain, the probabilities of the
LH$\to$CB and SO$\to$CB transitions are determined by the partial fractions of the
$| X \rangle$, $| Y \rangle$, and $| Z \rangle$ states in the initial state, although
these dependences are not exactly linear. The probability of transitions in the
$\pi$ polarization correlates with the fraction of $| Z \rangle$ states in the wave
function, and that for transitions in the $\sigma$ polarization correlates with the
fraction of $| X \pm iY \rangle$ states in the wave function. It is interesting that
despite strong changes in partial fractions of $|X \pm iY \rangle$ and $|Z \rangle$
states in the wave functions, the total intensity of the LH+SO$\to$CB transitions as well
as of the HH$\to$CB one changes by no more than 3.5--4\% over the entire strain range.
These are only the intensities of individual components that are changed.

\section{Conclusions}

An approach to analyzing the spinor wave functions that appear in the electronic
structure calculations when taking the spin-orbit interaction into account is developed.
It is based on the projection analysis of angular parts of wave functions onto
irreducible representations of the point group and analysis of the evolution of
the energy levels upon the adiabatic turning on the spin-orbit interaction. This
approach makes it possible to determine the genesis of the spinor wave functions,
calculate the fractions of different contributions to the mixing of these states,
and thus unambiguously identify electronic states. It seems promising to further use
this approach to identify electronic states in low-dimensional systems such as
superlattices, nanoplatelets, and nanoheterostructures, in which the valence band
states are strongly mixed, and their unambiguous identification is complicated by
the appearance of a large number of minibands as a result of the size quantization.

\begin{acknowledgments}
This work was partly supported by the Russian Science Foundation grant No.~22-13-00101.
\end{acknowledgments}

% Create the reference section using BibTeX:
%\bibliography{all}

\begin{thebibliography}{38}%
\makeatletter
\providecommand \@ifxundefined [1]{%
 \@ifx{#1\undefined}
}%
\providecommand \@ifnum [1]{%
 \ifnum #1\expandafter \@firstoftwo
 \else \expandafter \@secondoftwo
 \fi
}%
\providecommand \@ifx [1]{%
 \ifx #1\expandafter \@firstoftwo
 \else \expandafter \@secondoftwo
 \fi
}%
\providecommand \natexlab [1]{#1}%
\providecommand \enquote  [1]{``#1''}%
\providecommand \bibnamefont  [1]{#1}%
\providecommand \bibfnamefont [1]{#1}%
\providecommand \citenamefont [1]{#1}%
\providecommand \href@noop [0]{\@secondoftwo}%
\providecommand \href [0]{\begingroup \@sanitize@url \@href}%
\providecommand \@href[1]{\@@startlink{#1}\@@href}%
\providecommand \@@href[1]{\endgroup#1\@@endlink}%
\providecommand \@sanitize@url [0]{\catcode `\\12\catcode `\$12\catcode
  `\&12\catcode `\#12\catcode `\^12\catcode `\_12\catcode `\%12\relax}%
\providecommand \@@startlink[1]{}%
\providecommand \@@endlink[0]{}%
\providecommand \url  [0]{\begingroup\@sanitize@url \@url }%
\providecommand \@url [1]{\endgroup\@href {#1}{\urlprefix }}%
\providecommand \urlprefix  [0]{URL }%
\providecommand \Eprint [0]{\href }%
\providecommand \doibase [0]{https://doi.org/}%
\providecommand \selectlanguage [0]{\@gobble}%
\providecommand \bibinfo  [0]{\@secondoftwo}%
\providecommand \bibfield  [0]{\@secondoftwo}%
\providecommand \translation [1]{[#1]}%
\providecommand \BibitemOpen [0]{}%
\providecommand \bibitemStop [0]{}%
\providecommand \bibitemNoStop [0]{.\EOS\space}%
\providecommand \EOS [0]{\spacefactor3000\relax}%
\providecommand \BibitemShut  [1]{\csname bibitem#1\endcsname}%
\let\auto@bib@innerbib\@empty
%</preamble>
\bibitem [{\citenamefont {Kane}(1966)}]{SemicondSemimetals.1.Ch3}%
  \BibitemOpen
  \bibfield  {author} {\bibinfo {author} {\bibfnamefont {E.~O.}\ \bibnamefont
  {Kane}},\ }\bibinfo {title} {The ${\bf k \cdot p}$ method},\ in\ \href
  {https://doi.org/10.1016/S0080-8784(08)62376-5} {\emph {\bibinfo {booktitle}
  {Semiconductors and Semimetals}}},\ Vol.~\bibinfo {volume} {1}\ (\bibinfo
  {publisher} {Academic Press},\ \bibinfo {address} {New York, London},\
  \bibinfo {year} {1966})\ Chap.~\bibinfo {chapter} {3}, pp.\ \bibinfo {pages}
  {75--100}\BibitemShut {NoStop}%
\bibitem [{\citenamefont {Lew Yan~Voon}\ and\ \citenamefont
  {Willatzen}(2009)}]{VoonWillatzen}%
  \BibitemOpen
  \bibfield  {author} {\bibinfo {author} {\bibfnamefont {L.~C.}\ \bibnamefont
  {Lew Yan~Voon}}\ and\ \bibinfo {author} {\bibfnamefont {M.}~\bibnamefont
  {Willatzen}},\ }\href@noop {} {\emph {\bibinfo {title} {The ${\bf k \cdot p}$
  Method. Electronic Properties of Semiconductors}}}\ (\bibinfo  {publisher}
  {Springer-Verlag Berlin Heidelberg},\ \bibinfo {year} {2009})\BibitemShut
  {NoStop}%
\bibitem [{\citenamefont {L{\"o}wdin}(1951)}]{JChemPhys.19.1396}%
  \BibitemOpen
  \bibfield  {author} {\bibinfo {author} {\bibfnamefont {P.-O.}\ \bibnamefont
  {L{\"o}wdin}},\ }\bibfield  {title} {\bibinfo {title} {A note on the
  quantum-mechanical perturbation theory},\ }\href
  {https://doi.org/10.1063/1.1748067} {\bibfield  {journal} {\bibinfo
  {journal} {J. Chem. Phys.}\ }\textbf {\bibinfo {volume} {19}},\ \bibinfo
  {pages} {1396} (\bibinfo {year} {1951})}\BibitemShut {NoStop}%
\bibitem [{\citenamefont {Wood}\ and\ \citenamefont
  {Zunger}(1996)}]{PhysRevB.53.7949}%
  \BibitemOpen
  \bibfield  {author} {\bibinfo {author} {\bibfnamefont {D.~M.}\ \bibnamefont
  {Wood}}\ and\ \bibinfo {author} {\bibfnamefont {A.}~\bibnamefont {Zunger}},\
  }\bibfield  {title} {\bibinfo {title} {Successes and failures of the ${\bf k}
  \cdot {\bf p}$ method: A direct assessment for GaAs/AlAs quantum
  structures},\ }\href {https://doi.org/10.1103/PhysRevB.53.7949} {\bibfield
  {journal} {\bibinfo  {journal} {Phys. Rev. B}\ }\textbf {\bibinfo {volume}
  {53}},\ \bibinfo {pages} {7949} (\bibinfo {year} {1996})}\BibitemShut
  {NoStop}%
\bibitem [{\citenamefont {Fu}\ \emph {et~al.}(1998)\citenamefont {Fu},
  \citenamefont {Wang},\ and\ \citenamefont {Zunger}}]{PhysRevB.57.9971}%
  \BibitemOpen
  \bibfield  {author} {\bibinfo {author} {\bibfnamefont {H.}~\bibnamefont
  {Fu}}, \bibinfo {author} {\bibfnamefont {L.-W.}\ \bibnamefont {Wang}},\ and\
  \bibinfo {author} {\bibfnamefont {A.}~\bibnamefont {Zunger}},\ }\bibfield
  {title} {\bibinfo {title} {Applicability of the ${\bf k} \cdot {\bf p}$
  method to the electronic structure of quantum dots},\ }\href
  {https://doi.org/10.1103/PhysRevB.57.9971} {\bibfield  {journal} {\bibinfo
  {journal} {Phys. Rev. B}\ }\textbf {\bibinfo {volume} {57}},\ \bibinfo
  {pages} {9971} (\bibinfo {year} {1998})}\BibitemShut {NoStop}%
\bibitem [{\citenamefont {Zunger}(1999)}]{ElectrochemSocProc.98-19.259}%
  \BibitemOpen
  \bibfield  {author} {\bibinfo {author} {\bibfnamefont {A.}~\bibnamefont
  {Zunger}},\ }\bibfield  {title} {\bibinfo {title} {How to describe the
  electronic structure of semiconductor quantum dots},\ }in\ \href@noop {}
  {\emph {\bibinfo {booktitle} {Quantum Confinement V: Nanostructures}}},\
  Vol.\ \bibinfo {volume} {98-19},\ \bibinfo {editor} {edited by\ \bibinfo
  {editor} {\bibfnamefont {M.}~\bibnamefont {Cahay}}, \bibinfo {editor}
  {\bibfnamefont {D.~J.}\ \bibnamefont {Lockwood}}, \bibinfo {editor}
  {\bibfnamefont {J.~P.}\ \bibnamefont {Leburton}},\ and\ \bibinfo {editor}
  {\bibfnamefont {S.}~\bibnamefont {Bandyopadhyay}}}\ (\bibinfo  {publisher}
  {The Electrochemical Society},\ \bibinfo {year} {1999}),\ pp.\ \bibinfo
  {pages} {259--269}\BibitemShut {NoStop}%
\bibitem [{\citenamefont {Wang}\ and\ \citenamefont
  {Zunger}(1998)}]{JPhysChemB.102.6449}%
  \BibitemOpen
  \bibfield  {author} {\bibinfo {author} {\bibfnamefont {L.-W.}\ \bibnamefont
  {Wang}}\ and\ \bibinfo {author} {\bibfnamefont {A.}~\bibnamefont {Zunger}},\
  }\bibfield  {title} {\bibinfo {title} {High-energy excitonic transitions in
  CdSe quantum dots},\ }\href {https://doi.org/10.1021/jp981018n} {\bibfield
  {journal} {\bibinfo  {journal} {J. Phys. Chem. B}\ }\textbf {\bibinfo
  {volume} {102}},\ \bibinfo {pages} {6449} (\bibinfo {year}
  {1998})}\BibitemShut {NoStop}%
\bibitem [{\citenamefont {Wang}\ and\ \citenamefont
  {Zunger}(1996)}]{PhysRevB.54.11417}%
  \BibitemOpen
  \bibfield  {author} {\bibinfo {author} {\bibfnamefont {L.-W.}\ \bibnamefont
  {Wang}}\ and\ \bibinfo {author} {\bibfnamefont {A.}~\bibnamefont {Zunger}},\
  }\bibfield  {title} {\bibinfo {title} {Pseudopotential-based multiband ${\bf
  k} \cdot {\bf p}$ method for $\sim$250000-atom nanostructure system},\ }\href
  {https://doi.org/10.1103/PhysRevB.54.11417} {\bibfield  {journal} {\bibinfo
  {journal} {Phys. Rev. B}\ }\textbf {\bibinfo {volume} {54}},\ \bibinfo
  {pages} {11417} (\bibinfo {year} {1996})}\BibitemShut {NoStop}%
\bibitem [{\citenamefont {Iraola}\ \emph {et~al.}(2022)\citenamefont {Iraola},
  \citenamefont {Ma{\~n}es}, \citenamefont {Bradlyn}, \citenamefont {Horton},
  \citenamefont {Neupert}, \citenamefont {Vergniory},\ and\ \citenamefont
  {Tsirkin}}]{ComputPhysCommun.272.108226}%
  \BibitemOpen
  \bibfield  {author} {\bibinfo {author} {\bibfnamefont {M.}~\bibnamefont
  {Iraola}}, \bibinfo {author} {\bibfnamefont {J.~L.}\ \bibnamefont
  {Ma{\~n}es}}, \bibinfo {author} {\bibfnamefont {B.}~\bibnamefont {Bradlyn}},
  \bibinfo {author} {\bibfnamefont {M.~K.}\ \bibnamefont {Horton}}, \bibinfo
  {author} {\bibfnamefont {T.}~\bibnamefont {Neupert}}, \bibinfo {author}
  {\bibfnamefont {M.~G.}\ \bibnamefont {Vergniory}},\ and\ \bibinfo {author}
  {\bibfnamefont {S.~S.}\ \bibnamefont {Tsirkin}},\ }\bibfield  {title}
  {\bibinfo {title} {\mbox{IrRep}: Symmetry eigenvalues and irreducible
  representations of \emph{ab initio} band structures},\ }\href
  {https://doi.org/10.1016/j.cpc.2021.108226} {\bibfield  {journal} {\bibinfo
  {journal} {Comput. Phys. Commun.}\ }\textbf {\bibinfo {volume} {272}},\
  \bibinfo {pages} {108226} (\bibinfo {year} {2022})}\BibitemShut {NoStop}%
\bibitem [{\citenamefont {Klimov}(2010)}]{NanocrystalQuantumDots}%
  \BibitemOpen
  \bibinfo {editor} {\bibfnamefont {V.~I.}\ \bibnamefont {Klimov}},\ ed.,\
  \href@noop {} {\emph {\bibinfo {title} {Nanocrystal Quantum Dots}}}\
  (\bibinfo  {publisher} {CRC Press, Taylor \& Francis Group},\ \bibinfo
  {address} {Boca Raton, London, New York},\ \bibinfo {year}
  {2010})\BibitemShut {NoStop}%
\bibitem [{\citenamefont {Bera}\ \emph {et~al.}(2010)\citenamefont {Bera},
  \citenamefont {Qian}, \citenamefont {Tseng},\ and\ \citenamefont
  {Holloway}}]{Materials.3.2260}%
  \BibitemOpen
  \bibfield  {author} {\bibinfo {author} {\bibfnamefont {D.}~\bibnamefont
  {Bera}}, \bibinfo {author} {\bibfnamefont {L.}~\bibnamefont {Qian}}, \bibinfo
  {author} {\bibfnamefont {T.-K.}\ \bibnamefont {Tseng}},\ and\ \bibinfo
  {author} {\bibfnamefont {P.~H.}\ \bibnamefont {Holloway}},\ }\bibfield
  {title} {\bibinfo {title} {Quantum dots and their multimodal applications: A
  review},\ }\href {https://doi.org/10.3390/ma3042260} {\bibfield  {journal}
  {\bibinfo  {journal} {Materials}\ }\textbf {\bibinfo {volume} {3}},\ \bibinfo
  {pages} {2260} (\bibinfo {year} {2010})}\BibitemShut {NoStop}%
\bibitem [{\citenamefont {Diroll}\ \emph {et~al.}(2023)\citenamefont {Diroll},
  \citenamefont {Guzelturk}, \citenamefont {Po}, \citenamefont {Dabard},
  \citenamefont {Fu}, \citenamefont {Makke}, \citenamefont {Lhuillier},\ and\
  \citenamefont {Ithurria}}]{ChemRev.2c00436}%
  \BibitemOpen
  \bibfield  {author} {\bibinfo {author} {\bibfnamefont {B.~T.}\ \bibnamefont
  {Diroll}}, \bibinfo {author} {\bibfnamefont {B.}~\bibnamefont {Guzelturk}},
  \bibinfo {author} {\bibfnamefont {H.}~\bibnamefont {Po}}, \bibinfo {author}
  {\bibfnamefont {C.}~\bibnamefont {Dabard}}, \bibinfo {author} {\bibfnamefont
  {N.}~\bibnamefont {Fu}}, \bibinfo {author} {\bibfnamefont {L.}~\bibnamefont
  {Makke}}, \bibinfo {author} {\bibfnamefont {E.}~\bibnamefont {Lhuillier}},\
  and\ \bibinfo {author} {\bibfnamefont {S.}~\bibnamefont {Ithurria}},\
  }\bibfield  {title} {\bibinfo {title} {2D II--VI semiconductor nanoplatelets:
  From material synthesis to optoelectronic integration},\ }\href
  {https://doi.org/10.1021/acs.chemrev.2c00436} {\bibfield  {journal} {\bibinfo
   {journal} {Chem. Rev.}\ }\textbf {\bibinfo {volume} {123}},\ \bibinfo
  {pages} {3543} (\bibinfo {year} {2023})}\BibitemShut {NoStop}%
\bibitem [{\citenamefont {Stukel}\ \emph {et~al.}(1969)\citenamefont {Stukel},
  \citenamefont {Euwema}, \citenamefont {Collins}, \citenamefont {Herman},\
  and\ \citenamefont {Kortum}}]{PhysRev.179.740}%
  \BibitemOpen
  \bibfield  {author} {\bibinfo {author} {\bibfnamefont {D.~J.}\ \bibnamefont
  {Stukel}}, \bibinfo {author} {\bibfnamefont {R.~N.}\ \bibnamefont {Euwema}},
  \bibinfo {author} {\bibfnamefont {T.~C.}\ \bibnamefont {Collins}}, \bibinfo
  {author} {\bibfnamefont {F.}~\bibnamefont {Herman}},\ and\ \bibinfo {author}
  {\bibfnamefont {R.~L.}\ \bibnamefont {Kortum}},\ }\bibfield  {title}
  {\bibinfo {title} {Self-consistent orthogonalized-plane-wave and empirically
  refined orthogonalized-plane-wave energy-band models for cubic ZnS, ZnSe,
  CdS, and CdSe},\ }\href {https://doi.org/10.1103/PhysRev.179.740} {\bibfield
  {journal} {\bibinfo  {journal} {Phys. Rev.}\ }\textbf {\bibinfo {volume}
  {179}},\ \bibinfo {pages} {740} (\bibinfo {year} {1969})}\BibitemShut
  {NoStop}%
\bibitem [{\citenamefont {Huang}\ and\ \citenamefont
  {Ching}(1993)}]{PhysRevB.47.9449}%
  \BibitemOpen
  \bibfield  {author} {\bibinfo {author} {\bibfnamefont {M.-Z.}\ \bibnamefont
  {Huang}}\ and\ \bibinfo {author} {\bibfnamefont {W.~Y.}\ \bibnamefont
  {Ching}},\ }\bibfield  {title} {\bibinfo {title} {Calculation of optical
  excitations in cubic semiconductors. I. Electronic structure and linear
  response},\ }\href {https://doi.org/10.1103/PhysRevB.47.9449} {\bibfield
  {journal} {\bibinfo  {journal} {Phys. Rev. B}\ }\textbf {\bibinfo {volume}
  {47}},\ \bibinfo {pages} {9449} (\bibinfo {year} {1993})}\BibitemShut
  {NoStop}%
\bibitem [{\citenamefont {Kim}\ \emph {et~al.}(1994)\citenamefont {Kim},
  \citenamefont {Klein}, \citenamefont {Ren}, \citenamefont {Chang},
  \citenamefont {Luo}, \citenamefont {Samarth},\ and\ \citenamefont
  {Furdyna}}]{PhysRevB.49.7262}%
  \BibitemOpen
  \bibfield  {author} {\bibinfo {author} {\bibfnamefont {Y.~D.}\ \bibnamefont
  {Kim}}, \bibinfo {author} {\bibfnamefont {M.~V.}\ \bibnamefont {Klein}},
  \bibinfo {author} {\bibfnamefont {S.~F.}\ \bibnamefont {Ren}}, \bibinfo
  {author} {\bibfnamefont {Y.~C.}\ \bibnamefont {Chang}}, \bibinfo {author}
  {\bibfnamefont {H.}~\bibnamefont {Luo}}, \bibinfo {author} {\bibfnamefont
  {N.}~\bibnamefont {Samarth}},\ and\ \bibinfo {author} {\bibfnamefont {J.~K.}\
  \bibnamefont {Furdyna}},\ }\bibfield  {title} {\bibinfo {title} {Optical
  properties of zinc-blende CdSe and Zn$_x$Cd$_{1-x}$Se films grown on GaAs},\
  }\href {https://doi.org/10.1103/PhysRevB.49.7262} {\bibfield  {journal}
  {\bibinfo  {journal} {Phys. Rev. B}\ }\textbf {\bibinfo {volume} {49}},\
  \bibinfo {pages} {7262} (\bibinfo {year} {1994})}\BibitemShut {NoStop}%
\bibitem [{\citenamefont {Reshak}(2006)}]{JChemPhys.124.104707}%
  \BibitemOpen
  \bibfield  {author} {\bibinfo {author} {\bibfnamefont {A.~H.}\ \bibnamefont
  {Reshak}},\ }\bibfield  {title} {\bibinfo {title} {Theoretical investigation
  of the electronic properties, and first and second harmonic generation for
  cadmium chalcogenide},\ }\href {https://doi.org/10.1063/1.2178801} {\bibfield
   {journal} {\bibinfo  {journal} {J. Chem. Phys.}\ }\textbf {\bibinfo {volume}
  {124}},\ \bibinfo {pages} {104707} (\bibinfo {year} {2006})}\BibitemShut
  {NoStop}%
\bibitem [{\citenamefont {Reshak}\ \emph {et~al.}(2011)\citenamefont {Reshak},
  \citenamefont {Kityk}, \citenamefont {Khenata},\ and\ \citenamefont
  {Auluck}}]{JAlloysComp.509.6737}%
  \BibitemOpen
  \bibfield  {author} {\bibinfo {author} {\bibfnamefont {A.~H.}\ \bibnamefont
  {Reshak}}, \bibinfo {author} {\bibfnamefont {I.~V.}\ \bibnamefont {Kityk}},
  \bibinfo {author} {\bibfnamefont {R.}~\bibnamefont {Khenata}},\ and\ \bibinfo
  {author} {\bibfnamefont {S.}~\bibnamefont {Auluck}},\ }\bibfield  {title}
  {\bibinfo {title} {Effect of increasing tellurium content on the electronic
  and optical properties of cadmium selenide telluride alloys
  CdSe$_{1-x}$Te$_x$: An ab initio study},\ }\href
  {https://doi.org/10.1016/j.jallcom.2011.03.029} {\bibfield  {journal}
  {\bibinfo  {journal} {J. Alloys Comp.}\ }\textbf {\bibinfo {volume} {509}},\
  \bibinfo {pages} {6737} (\bibinfo {year} {2011})}\BibitemShut {NoStop}%
\bibitem [{\citenamefont {Ithurria}\ \emph {et~al.}(2011)\citenamefont
  {Ithurria}, \citenamefont {Tessier}, \citenamefont {Mahler}, \citenamefont
  {Lobo}, \citenamefont {Dubertret},\ and\ \citenamefont
  {Efros}}]{NatureMater.10.936}%
  \BibitemOpen
  \bibfield  {author} {\bibinfo {author} {\bibfnamefont {S.}~\bibnamefont
  {Ithurria}}, \bibinfo {author} {\bibfnamefont {M.~D.}\ \bibnamefont
  {Tessier}}, \bibinfo {author} {\bibfnamefont {B.}~\bibnamefont {Mahler}},
  \bibinfo {author} {\bibfnamefont {R.~P. S.~M.}\ \bibnamefont {Lobo}},
  \bibinfo {author} {\bibfnamefont {B.}~\bibnamefont {Dubertret}},\ and\
  \bibinfo {author} {\bibfnamefont {A.~L.}\ \bibnamefont {Efros}},\ }\bibfield
  {title} {\bibinfo {title} {Colloidal nanoplatelets with two-dimensional
  electronic structure},\ }\href {https://doi.org/10.1038/nmat3145} {\bibfield
  {journal} {\bibinfo  {journal} {Nat. Mater.}\ }\textbf {\bibinfo {volume}
  {10}},\ \bibinfo {pages} {936} (\bibinfo {year} {2011})}\BibitemShut
  {NoStop}%
\bibitem [{\citenamefont {Vasiliev}\ \emph {et~al.}(2017)\citenamefont
  {Vasiliev}, \citenamefont {Lebedev}, \citenamefont {Lazareva}, \citenamefont
  {Shlenskaya}, \citenamefont {Zaytsev}, \citenamefont {Vitukhnovsky},
  \citenamefont {Yao},\ and\ \citenamefont {Sakoda}}]{PhysRevB.95.165414}%
  \BibitemOpen
  \bibfield  {author} {\bibinfo {author} {\bibfnamefont {R.~B.}\ \bibnamefont
  {Vasiliev}}, \bibinfo {author} {\bibfnamefont {A.~I.}\ \bibnamefont
  {Lebedev}}, \bibinfo {author} {\bibfnamefont {E.~P.}\ \bibnamefont
  {Lazareva}}, \bibinfo {author} {\bibfnamefont {N.~N.}\ \bibnamefont
  {Shlenskaya}}, \bibinfo {author} {\bibfnamefont {V.~B.}\ \bibnamefont
  {Zaytsev}}, \bibinfo {author} {\bibfnamefont {A.~G.}\ \bibnamefont
  {Vitukhnovsky}}, \bibinfo {author} {\bibfnamefont {Y.}~\bibnamefont {Yao}},\
  and\ \bibinfo {author} {\bibfnamefont {K.}~\bibnamefont {Sakoda}},\
  }\bibfield  {title} {\bibinfo {title} {High-energy exciton transitions in
  quasi-two-dimensional cadmium chalcogenide nanoplatelets},\ }\href
  {https://doi.org/10.1103/PhysRevB.95.165414} {\bibfield  {journal} {\bibinfo
  {journal} {Phys. Rev. B}\ }\textbf {\bibinfo {volume} {95}},\ \bibinfo
  {pages} {165414} (\bibinfo {year} {2017})}\BibitemShut {NoStop}%
\bibitem [{\citenamefont {Bose}\ \emph {et~al.}(2016)\citenamefont {Bose},
  \citenamefont {Song}, \citenamefont {Fan},\ and\ \citenamefont
  {Zhang}}]{JApplPhys.119.143107}%
  \BibitemOpen
  \bibfield  {author} {\bibinfo {author} {\bibfnamefont {S.}~\bibnamefont
  {Bose}}, \bibinfo {author} {\bibfnamefont {Z.}~\bibnamefont {Song}}, \bibinfo
  {author} {\bibfnamefont {W.~J.}\ \bibnamefont {Fan}},\ and\ \bibinfo {author}
  {\bibfnamefont {D.~H.}\ \bibnamefont {Zhang}},\ }\bibfield  {title} {\bibinfo
  {title} {Effect of lateral size and thickness on the electronic structure and
  optical properties of quasi two-dimensional CdSe and CdS nanoplatelets},\
  }\href {https://doi.org/10.1063/1.4945993} {\bibfield  {journal} {\bibinfo
  {journal} {J. Appl. Phys.}\ }\textbf {\bibinfo {volume} {119}},\ \bibinfo
  {pages} {143107} (\bibinfo {year} {2016})}\BibitemShut {NoStop}%
\bibitem [{\citenamefont {Gonze}\ \emph {et~al.}(2009)\citenamefont {Gonze},
  \citenamefont {Amadon}, \citenamefont {Anglade}, \citenamefont {Beuken},
  \citenamefont {Bottin}, \citenamefont {Boulanger}, \citenamefont {Bruneval},
  \citenamefont {Caliste}, \citenamefont {Caracas}, \citenamefont {C\^ot\'e},
  \citenamefont {Deutsch}, \citenamefont {Genovese}, \citenamefont {Ghosez},
  \citenamefont {Giantomassi}, \citenamefont {Goedecker}, \citenamefont
  {Hamann}, \citenamefont {Hermet}, \citenamefont {Jollet}, \citenamefont
  {Jomard}, \citenamefont {Leroux}, \citenamefont {Mancini}, \citenamefont
  {Mazevet}, \citenamefont {Oliveira}, \citenamefont {Onida}, \citenamefont
  {Pouillon}, \citenamefont {Rangel}, \citenamefont {Rignanese}, \citenamefont
  {Sangalli}, \citenamefont {Shaltaf}, \citenamefont {Torrent}, \citenamefont
  {Verstraete}, \citenamefont {Zerah},\ and\ \citenamefont
  {Zwanziger}}]{abinit3}%
  \BibitemOpen
  \bibfield  {author} {\bibinfo {author} {\bibfnamefont {X.}~\bibnamefont
  {Gonze}}, \bibinfo {author} {\bibfnamefont {B.}~\bibnamefont {Amadon}},
  \bibinfo {author} {\bibfnamefont {P.-M.}\ \bibnamefont {Anglade}}, \bibinfo
  {author} {\bibfnamefont {J.-M.}\ \bibnamefont {Beuken}}, \bibinfo {author}
  {\bibfnamefont {F.}~\bibnamefont {Bottin}}, \bibinfo {author} {\bibfnamefont
  {P.}~\bibnamefont {Boulanger}}, \bibinfo {author} {\bibfnamefont
  {F.}~\bibnamefont {Bruneval}}, \bibinfo {author} {\bibfnamefont
  {D.}~\bibnamefont {Caliste}}, \bibinfo {author} {\bibfnamefont
  {R.}~\bibnamefont {Caracas}}, \bibinfo {author} {\bibfnamefont
  {M.}~\bibnamefont {C\^ot\'e}}, \bibinfo {author} {\bibfnamefont
  {T.}~\bibnamefont {Deutsch}}, \bibinfo {author} {\bibfnamefont
  {L.}~\bibnamefont {Genovese}}, \bibinfo {author} {\bibfnamefont
  {P.}~\bibnamefont {Ghosez}}, \bibinfo {author} {\bibfnamefont
  {M.}~\bibnamefont {Giantomassi}}, \bibinfo {author} {\bibfnamefont
  {S.}~\bibnamefont {Goedecker}}, \bibinfo {author} {\bibfnamefont {D.~R.}\
  \bibnamefont {Hamann}}, \bibinfo {author} {\bibfnamefont {P.}~\bibnamefont
  {Hermet}}, \bibinfo {author} {\bibfnamefont {F.}~\bibnamefont {Jollet}},
  \bibinfo {author} {\bibfnamefont {G.}~\bibnamefont {Jomard}}, \bibinfo
  {author} {\bibfnamefont {S.}~\bibnamefont {Leroux}}, \bibinfo {author}
  {\bibfnamefont {M.}~\bibnamefont {Mancini}}, \bibinfo {author} {\bibfnamefont
  {S.}~\bibnamefont {Mazevet}}, \bibinfo {author} {\bibfnamefont {M.~J.~T.}\
  \bibnamefont {Oliveira}}, \bibinfo {author} {\bibfnamefont {G.}~\bibnamefont
  {Onida}}, \bibinfo {author} {\bibfnamefont {Y.}~\bibnamefont {Pouillon}},
  \bibinfo {author} {\bibfnamefont {T.}~\bibnamefont {Rangel}}, \bibinfo
  {author} {\bibfnamefont {G.-M.}\ \bibnamefont {Rignanese}}, \bibinfo {author}
  {\bibfnamefont {D.}~\bibnamefont {Sangalli}}, \bibinfo {author}
  {\bibfnamefont {R.}~\bibnamefont {Shaltaf}}, \bibinfo {author} {\bibfnamefont
  {M.}~\bibnamefont {Torrent}}, \bibinfo {author} {\bibfnamefont {M.~J.}\
  \bibnamefont {Verstraete}}, \bibinfo {author} {\bibfnamefont
  {G.}~\bibnamefont {Zerah}},\ and\ \bibinfo {author} {\bibfnamefont {J.~W.}\
  \bibnamefont {Zwanziger}},\ }\bibfield  {title} {\bibinfo {title} {ABINIT:
  First-principles approach to material and nanosystem properties},\ }\href
  {https://doi.org/10.1016/j.cpc.2009.07.007} {\bibfield  {journal} {\bibinfo
  {journal} {Comput. Phys. Commun.}\ }\textbf {\bibinfo {volume} {180}},\
  \bibinfo {pages} {2582} (\bibinfo {year} {2009})}\BibitemShut {NoStop}%
\bibitem [{\citenamefont {Garrity}\ \emph {et~al.}(2014)\citenamefont
  {Garrity}, \citenamefont {Bennett}, \citenamefont {Rabe},\ and\ \citenamefont
  {Vanderbilt}}]{ComputMaterSci.81.446}%
  \BibitemOpen
  \bibfield  {author} {\bibinfo {author} {\bibfnamefont {K.~F.}\ \bibnamefont
  {Garrity}}, \bibinfo {author} {\bibfnamefont {J.~W.}\ \bibnamefont
  {Bennett}}, \bibinfo {author} {\bibfnamefont {K.~M.}\ \bibnamefont {Rabe}},\
  and\ \bibinfo {author} {\bibfnamefont {D.}~\bibnamefont {Vanderbilt}},\
  }\bibfield  {title} {\bibinfo {title} {Pseudopotentials for high throughput
  DFT calculations},\ }\href {https://doi.org/10.1016/j.commatsci.2013.08.053}
  {\bibfield  {journal} {\bibinfo  {journal} {Comput. Mater. Sci.}\ }\textbf
  {\bibinfo {volume} {81}},\ \bibinfo {pages} {446} (\bibinfo {year}
  {2014})}\BibitemShut {NoStop}%
\bibitem [{\citenamefont {Hartwigsen}\ \emph {et~al.}(1998)\citenamefont
  {Hartwigsen}, \citenamefont {Goedecker},\ and\ \citenamefont
  {Hutter}}]{PhysRevB.58.3641}%
  \BibitemOpen
  \bibfield  {author} {\bibinfo {author} {\bibfnamefont {C.}~\bibnamefont
  {Hartwigsen}}, \bibinfo {author} {\bibfnamefont {S.}~\bibnamefont
  {Goedecker}},\ and\ \bibinfo {author} {\bibfnamefont {J.}~\bibnamefont
  {Hutter}},\ }\bibfield  {title} {\bibinfo {title} {Relativistic separable
  dual-space gaussian pseudopotentials from H to Rn},\ }\href
  {https://doi.org/10.1103/PhysRevB.58.3641} {\bibfield  {journal} {\bibinfo
  {journal} {Phys. Rev. B}\ }\textbf {\bibinfo {volume} {58}},\ \bibinfo
  {pages} {3641} (\bibinfo {year} {1998})}\BibitemShut {NoStop}%
\bibitem [{\citenamefont {Hamann}(2013)}]{PhysRevB.88.085117}%
  \BibitemOpen
  \bibfield  {author} {\bibinfo {author} {\bibfnamefont {D.~R.}\ \bibnamefont
  {Hamann}},\ }\bibfield  {title} {\bibinfo {title} {Optimized norm-conserving
  Vanderbilt pseudopotentials},\ }\href
  {https://doi.org/10.1103/PhysRevB.88.085117} {\bibfield  {journal} {\bibinfo
  {journal} {Phys. Rev. B}\ }\textbf {\bibinfo {volume} {88}},\ \bibinfo
  {pages} {085117} (\bibinfo {year} {2013})}\BibitemShut {NoStop}%
\bibitem [{\citenamefont {Ricci}\ \emph {et~al.}(2019)\citenamefont {Ricci},
  \citenamefont {Prokhorenko}, \citenamefont {Torrent}, \citenamefont
  {Verstraete},\ and\ \citenamefont {Bousquet}}]{PhysRevB.99.184404}%
  \BibitemOpen
  \bibfield  {author} {\bibinfo {author} {\bibfnamefont {F.}~\bibnamefont
  {Ricci}}, \bibinfo {author} {\bibfnamefont {S.}~\bibnamefont {Prokhorenko}},
  \bibinfo {author} {\bibfnamefont {M.}~\bibnamefont {Torrent}}, \bibinfo
  {author} {\bibfnamefont {M.~J.}\ \bibnamefont {Verstraete}},\ and\ \bibinfo
  {author} {\bibfnamefont {E.}~\bibnamefont {Bousquet}},\ }\bibfield  {title}
  {\bibinfo {title} {Density functional perturbation theory within noncollinear
  magnetism},\ }\href {https://doi.org/10.1103/PhysRevB.99.184404} {\bibfield
  {journal} {\bibinfo  {journal} {Phys. Rev. B}\ }\textbf {\bibinfo {volume}
  {99}},\ \bibinfo {pages} {184404} (\bibinfo {year} {2019})}\BibitemShut
  {NoStop}%
\bibitem [{\citenamefont {Altmann}\ and\ \citenamefont
  {Herzig}(1994)}]{AltmannHerzig}%
  \BibitemOpen
  \bibfield  {author} {\bibinfo {author} {\bibfnamefont {S.~L.}\ \bibnamefont
  {Altmann}}\ and\ \bibinfo {author} {\bibfnamefont {P.}~\bibnamefont
  {Herzig}},\ }\href@noop {} {\emph {\bibinfo {title} {Point-Group Theory
  Tables}}}\ (\bibinfo  {publisher} {Clarendon Press},\ \bibinfo {year}
  {1994})\BibitemShut {NoStop}%
\bibitem [{\citenamefont {Bir}\ and\ \citenamefont
  {Pikus}(1974)}]{BirPikus1972-transl}%
  \BibitemOpen
  \bibfield  {author} {\bibinfo {author} {\bibfnamefont {G.~L.}\ \bibnamefont
  {Bir}}\ and\ \bibinfo {author} {\bibfnamefont {G.~E.}\ \bibnamefont
  {Pikus}},\ }\href@noop {} {\emph {\bibinfo {title} {Symmetry and
  Strain-induced Effects in Semiconductors}}}\ (\bibinfo  {publisher} {John
  Wiley \& Sons},\ \bibinfo {address} {New York},\ \bibinfo {year}
  {1974})\BibitemShut {NoStop}%
\bibitem [{\citenamefont {Koster}\ \emph {et~al.}(1963)\citenamefont {Koster},
  \citenamefont {Dimmock}, \citenamefont {Wheeler},\ and\ \citenamefont
  {Statz}}]{Koster1963}%
  \BibitemOpen
  \bibfield  {author} {\bibinfo {author} {\bibfnamefont {G.~F.}\ \bibnamefont
  {Koster}}, \bibinfo {author} {\bibfnamefont {J.~O.}\ \bibnamefont {Dimmock}},
  \bibinfo {author} {\bibfnamefont {R.~G.}\ \bibnamefont {Wheeler}},\ and\
  \bibinfo {author} {\bibfnamefont {H.}~\bibnamefont {Statz}},\ }\href@noop {}
  {\emph {\bibinfo {title} {Properties of The Thirty-Two Point Groups}}}\
  (\bibinfo  {publisher} {MIT Press, Cambridge MA},\ \bibinfo {year}
  {1963})\BibitemShut {NoStop}%
\bibitem [{\citenamefont {Luttinger}\ and\ \citenamefont
  {Kohn}(1955)}]{PhysRev.97.869}%
  \BibitemOpen
  \bibfield  {author} {\bibinfo {author} {\bibfnamefont {J.~M.}\ \bibnamefont
  {Luttinger}}\ and\ \bibinfo {author} {\bibfnamefont {W.}~\bibnamefont
  {Kohn}},\ }\bibfield  {title} {\bibinfo {title} {Motion of electrons and
  holes in perturbed periodic fields},\ }\href
  {https://doi.org/10.1103/PhysRev.97.869} {\bibfield  {journal} {\bibinfo
  {journal} {Phys. Rev.}\ }\textbf {\bibinfo {volume} {97}},\ \bibinfo {pages}
  {869} (\bibinfo {year} {1955})}\BibitemShut {NoStop}%
\bibitem [{\citenamefont {Dresselhaus}(1955)}]{PhysRev.100.580}%
  \BibitemOpen
  \bibfield  {author} {\bibinfo {author} {\bibfnamefont {G.}~\bibnamefont
  {Dresselhaus}},\ }\bibfield  {title} {\bibinfo {title} {Spin-orbit coupling
  effects in zinc blende structures},\ }\href
  {https://doi.org/10.1103/PhysRev.100.580} {\bibfield  {journal} {\bibinfo
  {journal} {Phys. Rev.}\ }\textbf {\bibinfo {volume} {100}},\ \bibinfo {pages}
  {580} (\bibinfo {year} {1955})}\BibitemShut {NoStop}%
\bibitem [{\citenamefont {Kane}(1957)}]{JPhysChemSolids.1.249}%
  \BibitemOpen
  \bibfield  {author} {\bibinfo {author} {\bibfnamefont {E.~O.}\ \bibnamefont
  {Kane}},\ }\bibfield  {title} {\bibinfo {title} {Band structure of indium
  antimonide},\ }\href {https://doi.org/10.1016/0022-3697(57)90013-6}
  {\bibfield  {journal} {\bibinfo  {journal} {J. Phys. Chem. Solids}\ }\textbf
  {\bibinfo {volume} {1}},\ \bibinfo {pages} {249} (\bibinfo {year}
  {1957})}\BibitemShut {NoStop}%
\bibitem [{\citenamefont {Chuang}(1995)}]{PhysicsOptoelectronicDevices}%
  \BibitemOpen
  \bibfield  {author} {\bibinfo {author} {\bibfnamefont {S.~L.}\ \bibnamefont
  {Chuang}},\ }\href@noop {} {\emph {\bibinfo {title} {Physics of
  Optoelectronic Devices}}}\ (\bibinfo  {publisher} {John Wiley \& Sons,
  Inc.},\ \bibinfo {year} {1995})\BibitemShut {NoStop}%
\bibitem [{\citenamefont {Jones}\ and\ \citenamefont
  {O'Reilly}(1993)}]{IEEEJQuantElectron.29.1344}%
  \BibitemOpen
  \bibfield  {author} {\bibinfo {author} {\bibfnamefont {G.}~\bibnamefont
  {Jones}}\ and\ \bibinfo {author} {\bibfnamefont {E.~P.}\ \bibnamefont
  {O'Reilly}},\ }\bibfield  {title} {\bibinfo {title} {Improved performance of
  long-wavelength strained bulk-like semiconductor lasers},\ }\href
  {https://doi.org/10.1109/3.236148} {\bibfield  {journal} {\bibinfo  {journal}
  {IEEE J. Quant. Electron.}\ }\textbf {\bibinfo {volume} {29}},\ \bibinfo
  {pages} {1344} (\bibinfo {year} {1993})}\BibitemShut {NoStop}%
\bibitem [{Note1()}]{Note1}%
  \BibitemOpen
  \bibinfo {note} {Due to the absence of an inversion center in crystals with
  the $D_{2d}^9$ space group, the spinor wave functions are described by two
  independent components at all points of the Brillouin zone except for the
  $\Gamma $, $X$, $M$ points and the points on the~$\Lambda $
  axis.}\BibitemShut {Stop}%
\bibitem [{\citenamefont {Blacha}\ \emph {et~al.}(1984)\citenamefont {Blacha},
  \citenamefont {Presting},\ and\ \citenamefont
  {Cardona}}]{PhysStatSolidiB.126.11}%
  \BibitemOpen
  \bibfield  {author} {\bibinfo {author} {\bibfnamefont {A.}~\bibnamefont
  {Blacha}}, \bibinfo {author} {\bibfnamefont {H.}~\bibnamefont {Presting}},\
  and\ \bibinfo {author} {\bibfnamefont {M.}~\bibnamefont {Cardona}},\
  }\bibfield  {title} {\bibinfo {title} {Deformation potentials of $k = 0$
  states in tetrahedral semiconductors},\ }\href
  {https://doi.org/10.1002/pssb.2221260102} {\bibfield  {journal} {\bibinfo
  {journal} {Phys. Status Solidi B}\ }\textbf {\bibinfo {volume} {126}},\
  \bibinfo {pages} {11} (\bibinfo {year} {1984})}\BibitemShut {NoStop}%
\bibitem [{\citenamefont {Dyakonov}\ and\ \citenamefont
  {Perel}(1984)}]{OpticalOrientation}%
  \BibitemOpen
  \bibfield  {author} {\bibinfo {author} {\bibfnamefont {M.~I.}\ \bibnamefont
  {Dyakonov}}\ and\ \bibinfo {author} {\bibfnamefont {V.~I.}\ \bibnamefont
  {Perel}},\ }\bibfield  {title} {\bibinfo {title} {Theory of optical spin
  orientation of electrons and niclei in semiconductors},\ }in\ \href@noop {}
  {\emph {\bibinfo {booktitle} {Optical Orientation}}},\ \bibinfo {editor}
  {edited by\ \bibinfo {editor} {\bibfnamefont {F.}~\bibnamefont {Meier}}\ and\
  \bibinfo {editor} {\bibfnamefont {B.~P.}\ \bibnamefont {Zakharchenya}}}\
  (\bibinfo  {publisher} {Elsevier Science Publisher B. V.},\ \bibinfo {year}
  {1984})\ Chap.~\bibinfo {chapter} {2}, pp.\ \bibinfo {pages}
  {11--71}\BibitemShut {NoStop}%
\bibitem [{Note2()}]{Note2}%
  \BibitemOpen
  \bibinfo {note} {To do this, we need a tetragonal strain; without it, the
  matrix elements in all polarizations are the same (cubic
  symmetry).}\BibitemShut {Stop}%
\bibitem [{Note3()}]{Note3}%
  \BibitemOpen
  \bibinfo {note} {In particular, this peculiarity results in appearance of
  forbidden transitions with $\Delta m_j = 2$ when calculating the HH$\to $CB
  matrix elements with these wave functions.}\BibitemShut {Stop}%
\end{thebibliography}

%apsrev4-2.bst 2019-01-14 (MD) hand-edited version of apsrev4-1.bst
%Control: key (0)
%Control: author (8) initials jnrlst
%Control: editor formatted (1) identically to author
%Control: production of article title (0) allowed
%Control: page (0) single
%Control: year (1) truncated
%Control: production of eprint (0) enabled
\providecommand{\BIBYu}{Yu}

\end{document}